# Meta-mimetism in Spatial Games

David Chavalarias[*] , Paul Bourgine

*Center for Research in Applied Epistemology (CREA), Ecole Polytechnique, Paris, France, www.crea.polytechnique.fr*

**Much attention has been given last years to the modelling of social systems, in economy as well as in anthropology or in political sciences. Among the phenomena considered as having a structuring power in social systems, imitation processes play a central role, completing and sometime competing with a more traditional economic approach based on the rational choice theory.**

**But the diversity of mimetic rules employed by modellers proves that the introduction of mimetic processes into formal models can't avoid the traditional problem of endogenization of all the choices, including the one of the mimetic laws. The aim of this article is to address this question starting from some studies about cognitive differences between humans and animals. This will lead us to propose a formal framework which has the advantage of endogenising mimetic processes and give a comprehensive description of a heterogeneous population structure at the behavioural level as well as at the level of the choices of mimetic laws.**

**In order to illustrate this point, we will apply this formalism to a spatial evolutionary game.**

---

[*] Correspondence and requests for materials should be addressed to David Chavalarias : chavalar@poly.polytechnique.fr

**The specificity of Human mimetic processes**

Scientists, to account for the extremely rich structures observed in human social systems, more and more often study mimetic processes. But this interest is not only due to the fact that mimetism is a component of human behaviour; after some theories[10], the sophistication of pre-human mimetic processes could have been the first step in hominization toward human social organisation as we know it.

In the literature of social systems modelling, two main processes of imitation have been defined. (1): in the traditional conception of *Homo oeconomicus*, some researchers considered *payoffs-biased imitation* i.e. imitation of the most successful agents in one's neighborhood[23]. (2): a growing number of contributions are attempts to introduce what is called *conformism*, in the study of social phenomena[5,6,11,25]. Here, *conformism* is the propensity of individuals to adopt some behaviour when it has already been adopted by some of their neighbors, the propensity being relative to the frequency of that behavior in the neighborhood[7]. To a lesser extent, other imitation processes have been studied, among which we can mention (3): *anti-conformism*, the propensity of an individual to adopt the behavior of the minority[3]. This list of imitation processes is far from exhaustive and we can already notice that even for *conformism* or *payoffs-biased imitation*, several technical definitions have been proposed, either deterministic or probabilistic[24]. On the other hand, it also possible to propose models including several imitations laws, as some authors already began to do it[8,14,18,19].

This raises an epistemological question for modellers. Which rule(s) for imitation should be considered depending on the social systems under study? Since the dynamics of the systems studied are very sensitive to the very definition of theses imitation rules, it is important to have a way to decide whether some rules are more suitable than others in the modelling of specific social phenomena. Moreover, since the distribution of these rules for imitation should be set by the system itself, it would be interesting to have some models with self-adaptive imitation rules distribution. To address this question, we will adopt in the following sections, a definition of rule for imitation close to what has been called *selective attention*, i.e. a mechanism that enables someone to select in her environment a subset of people that she will learn from. This operation of selection is exactly the role of a mimetic rule. To be more precise we will consider to be an imitation rule any mechanism which, given the neighbourhood's properties of an agent, her the personal state, and a cultural modifiable feature, points as output toward a particular member of the population, or a subgroup, from which the agent will be influenced to modify this modifiable feature (fig. 1). Typically, *conformism* – imitation of the agents with the most common features – or *payoffs-biased imitation* – imitation of the most successful agents - are such kinds of rules. By modifiable features we mean some features the agent can change by its own will on a small time scales (within a day), like a cooperative vs. defective behavior, the colours of the clothes she wears, etc. We will oppose them to features changing on larger time scales (months, years) and depending on a global dynamics, like the social position of the agents, her cumulated payoffs, her reputation, etc. The information included in the neighbourhood properties can be very rich and include distributions of other modifiable or not modifiable features.

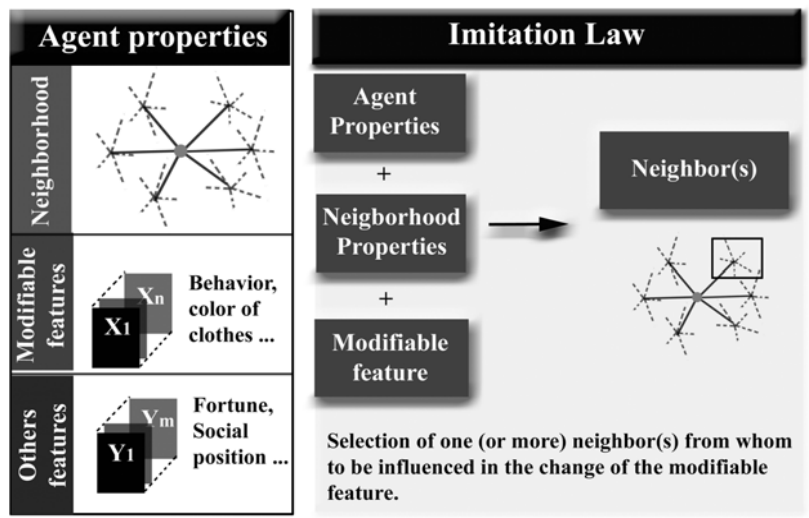

Figure 1: Imitation rule is any mechanism which, given a social structure, your personal state and a cultural modifiable feature, points as output toward a particular member, or a subgroup, of the population from which you will be influenced to modify this feature.

Since the specificity of human mimetism seems to be a central point in the understanding of human social systems, the first thing to do is to try to catch some differences between human and animal imitation processes. What can be surprising for a non-specialist looking at recent researches in ethology is that a lot of features, which at first glance can be thought to characterize human behaviours, can be found in some particular species (syntactic communication, dialogue, sharing, teaching, punishment etc.)[22]. But some other features definitely seem to belong only to humans. Among them[21]: heterotechnic cooperation (cooperation of several individuals with distinct skills in order to achieve a given task), the use of polylithes[26] (tools composed by several pieces assembled together by fastenings), individual learning of collective use of tools, or the narrative capacity. All these features have in common some very human specific abilities. First, they require a reflexive attitude from agents. Agents have to think themselves as a part of the surrounding world, which can be subject to same operations that she practices on external entities (non-self). Secondly, they require the capacity to jump from a cognitive level to its meta-level, i.e. a higher cognitive level where elements of the lower became objects submitted to transformation or recombination. The specificity of human meta-cognitive capacities is highlighted by several laboratory experiments showing that animals, contrary to humans, have very limited capacities to build meta-representations[27].

As for mimetic processes, reflexivity and meta-representation capacities introduced a big gap between animal and human mimetic processes[10]. If - as Eric Gans says[12] -, *"prehuman imitation is non-reflexive; the subject has no knowledge of itself as a self imitating another"*, humans are conscious that they are mimetic entities and in some extent, can monitor their mimetic behaviour[27]. From a formal point of view, this means that there is a loop between the agent's actions and the mimetic dynamics. It is this loop that we will try to make more explicit in the following.

**Modelling meta-choices**

From the definition we took for imitation rules and the remarks of last section, we can now see the effects of human specific abilities on mimetic processes. When imitation processes become reflexive and subject to meta-representations:

i) Rules for imitation became modifiable features by way of meta-rules application (figure 2-b-2),

ii) An imitation rule can operate reflexively controlling its own expression (figure 2-c).

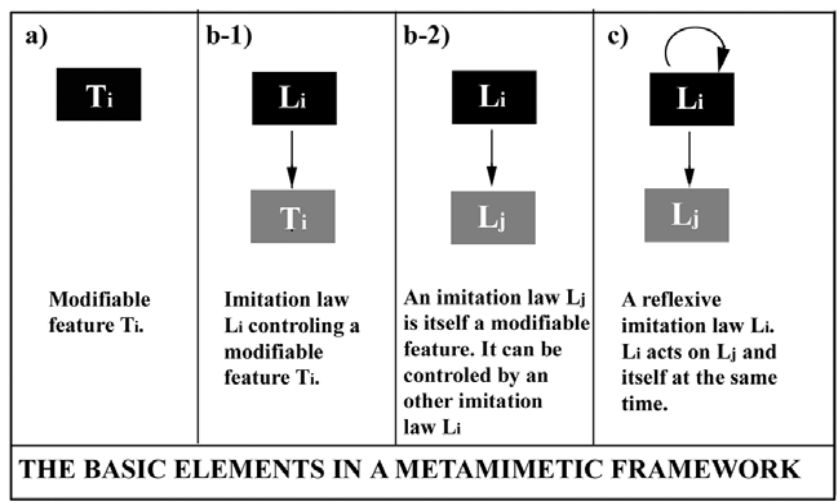

Figure 2

Assumption i) says that if an imitation rules is a modifiable feature, it can be controlled by an other imitation rule, its meta-rule. Human mimetic processes can then be composed of several levels, the upper controlling the lower (fig 2-b-2). For a given modifiable feature, we can associate the chain of mimetic rules that control its expression. But this chain of mimetic rules ought to be finite since humans have only limited cognitive capacities. Two solutions arise for the last layer setup. First, it could be fixed for all lifetime by a biological evolutionary process. Secondly, it could evolve through social influence mechanism during the individual lifetime. If we retain the last solution, then the last layer ought to be composed by reflexive imitation rules. These special self-applying rules for imitation are then the keys for endogenization. In a sense, human imitation rules have the particularity to be potentially their own meta-rules. On this basis we will propose some cognitive hypotheses:

Cognitive hypotheses:

1. Human mimetic systems can be composed of several meta-levels but their number is finite.

2. In human cognitive systems, at all levels, rules for imitation are modifiable features.

3. The last meta-level in a mimetic system loops on itself.

As we saw it, $(1)+(2) \Rightarrow (3)$. These hypothesis and the kinds of mimetic structures it implies enable us to adopt a new approach of the modelling of mimetic processes. With the introduction of these specific human capacities in formal models, a new kind of dynamics appears to modellers: dynamics on imitation rules and their meta-rules. Their study should reveal some particular patterns similar to what can be observed in the organization of human societies, or at least, give us some powerful analogies to think human social systems. This also introduises a change in the methodology for formal studies. Since the set of mimetic rules has its own internal dynamics, the problem of the modeller is now A) to decide which mimetic structures agents can have (the number of meta-levels), and which set of imitation rules is relevant for the problem she tries to model; B) which attractors of the meta-mimetic dynamics ought to be considered.

The answer to the first question depends of the cognitive capacities of the entities under study. Since imitation processes are comparison between several agents in order to select the one you will imitate, imitation supposes that these agents can be compared on some particular scale (the amount of their fortune, their reputation, the size of the group they belong to, etc.). The dimensions on which imitation processes will be able to operate will then depend on the cognitive level of the modelised population: the number of salient percepts its members will be able to extract from the environment and the way they will be able to make comparison between these percepts. The technological level of populations studied will also have a great influence on mimetic dynamics: technology, through the introduction of new aggregated data available for the agents (pools, surveys, etc.) and changes introduced in the topology of their neighborhood (radio, TV, e-mail, Internet, etc.), allows for new imitation rules definition and changes the means of diffusion of the existing ones.

To answer the second question, we have to look closer at the mimetic dynamics themselves. The first studies about these systems indicate that the number of attractors of these dynamics is very limited compared to the complexity of the model, making the choice of an attractor reasonable.

**An application of Meta-mimetism to a simple spatial game**

In this section we will apply the meta-mimetism formalism to the spatial game proposed by Novak and May in 1992[23]. Although there have been a lot of developments based on this model[24] and a wider number of models on emergence of cooperation, taking into account diverse social or cognitive structures[1,2,9,15,16,25], its simplicity makes it a good example to start with. The aim of this section is not to present a new model for emergence of cooperation, but to give a concrete example of application of meta-mimetism. Novak and May considered agents on a two dimensional toric lattice playing a two-players game each round with each agent of their neighbourhood. Players can only adopt one of two strategies: always cooperate, or always defect. The neighbourhood of a player is composed by the eight adjacent cells and, in case of models with self-interaction, its own cell. Following the structure of their article, we will present results concerning models with self-interaction. Similar results have been obtained without self-interaction[i]. In their model, when two agents play together, they receive a payoff of *1* if both cooperate (*C*), and *0* if both defect (*D*). In case they play different strategies, the one who defects gets a payoff of *b>1* and the other gets *0*. At the end of each round all agents change their behaviour to the behaviour of the most successful agent in their neighborhood. In case they are one of the most successful, they keep their strategy. It is a pure payoffs-biased mechanism.

The main results of Nowak and May was that this spatial game generates a spatial structure within populations with coexistence of both *C* and *D* types for some particular value of *b (1.8<b<2)*. The structures of theses spatial patterns vary also with *b* values from static fractal states to chaotic evolving structures. Outside the range *[1.8,2]*, the behavior of agents is uniform, all *D* or all *C* across the population.

To extend this model in the meta-mimetism framework, we first have to choose a structure for the agents. For this first application, we will consider the simplest closed architecture: two levels, the modifiable features (*D* or *C*) and the meta-level with reflexive mimetic rules (figure 3). Although human cognitive imitation processes are probably much more complex, this structure will give us a good approach of these meta-dynamics.

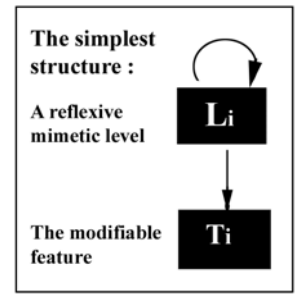

**Figure 3**

As for imitation rules, we will extend the set of available imitation rules to the set of rules mentioned in first section:

1. *Maxi*: the original *payoffs-biased* rule, "copy the most successful agent in your neighborhood".

2. *Lmaj*: the majority rule, "copy the majority" (*conformism*)

3. *Lmno*: the minority rule, "copy the minority" (*anti-conformism*)

    We will add to this set a stochastic rule

4. *PDR* (Proportional Density Rule): the stochastic conformist law: "copy a feature proportionally to its frequency in your neighborhood"[1].

This rule is obtained for example if you consider an agent that imitates at random in her neighbourhood. The algorithm used for the simulations presented here can be found in the methodology paragraph.

---

[1] For all these rules: in case of equality, agents choose at random. But they change anything if they are one of the neighbors selected in the mimetic process.

To begin with, let us look at the behavior of these systems for *b=1.9* and for a uniform initial distribution of imitation rules and actions. The parameters are corresponding to a chaotic regime in the original model. The statistics of these simulations are shown in figure 4. The first observation is that the system reaches very quickly its unique attractor (*20* periods), which is mostly static (for a *10 000* agents population, about a hundred of oscillators at the level of imitation rules, less than a dozen at the behavioral level). Moreover the path to this attractor is mostly the same along different simulations with a very low variance on the imitation rules distribution as well as on the distribution of behaviors. The distributions shown here are stable on the long run. The asymptotic state is heterogeneous at the imitation's rules level as well as at the action's rules level.

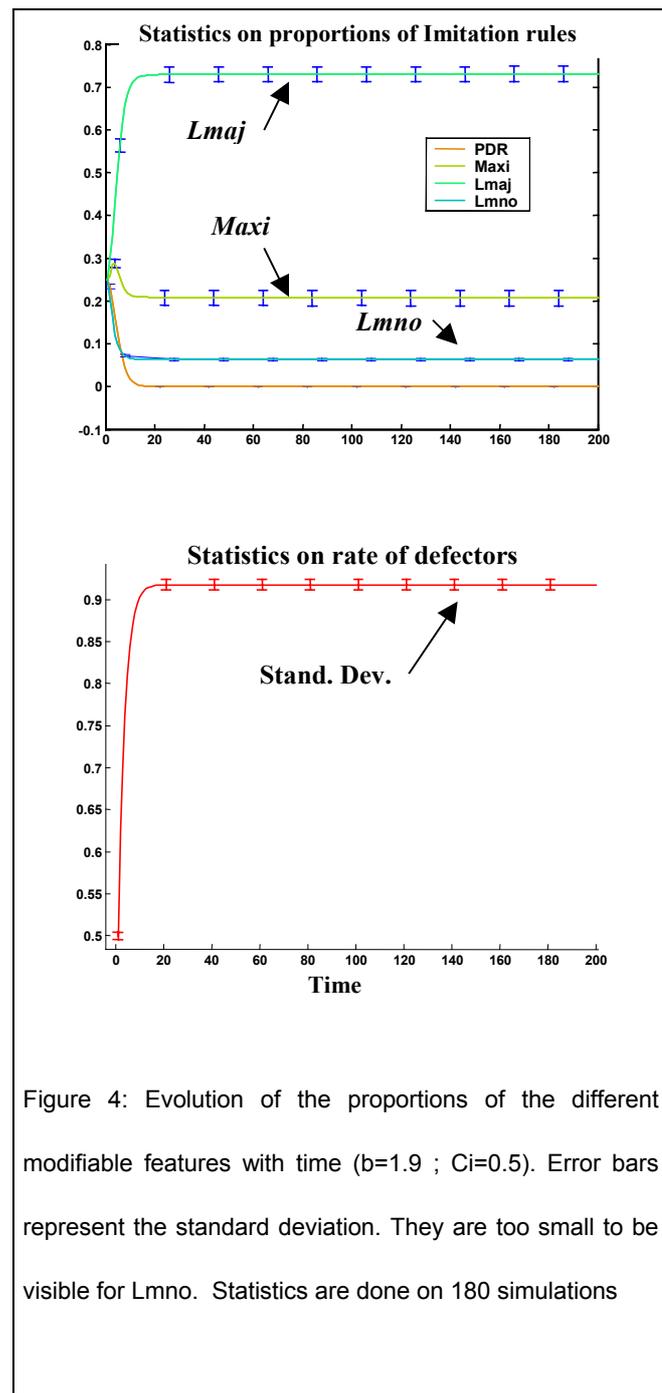

Figure 4: Evolution of the proportions of the different modifiable features with time (b=1.9 ; Ci=0.5). Error bars represent the standard deviation. They are too small to be visible for Lmno. Statistics are done on 180 simulations

On the initial four rules, only three are left, *PDR* having completely disappeared. The most common rule in the population is *Lmaj (70%)* followed by *Maxi (20%)*.

If we now look at the spatial distribution of the agent's modifiable features (figure 5), we can notice a very strong structuration of the population. If *Lmaj* agents fill the majority of the space, *Maxi* agents tend to form clusters while *Lmno* agents are scattered on all the territory. As for behaviours, *Lmaj* cooperators tend to form small groups while *Lmno* cooperators are isolated.

We did the precedent study for $1.2 \leq b \leq 2.5$ and for the initial rate of cooperators *Ci* ranging between *0.1* and *0.9*. A detailed presentation of simulation results can be found on the [web][i]. From this study, we can conclude that the ranking of the proportions of the different rules for imitation is the same along all the parameter values studied. The variations of these proportions (between *0.05* and *0.33* for Max; *0.65* and *0.89* for *Lmaj*; *0.05* and *0.07* for *Lmno*) show that the different attractors are qualitatively similar: clusters of *Maxi* agents in ocean of Lmaj agents with scattered *Lmno* agents. *PDR* agents are never present at the attractor. From the four selected laws initially available, only three are relevant in this game: *Maxi, Lmaj* and *Lmno*.

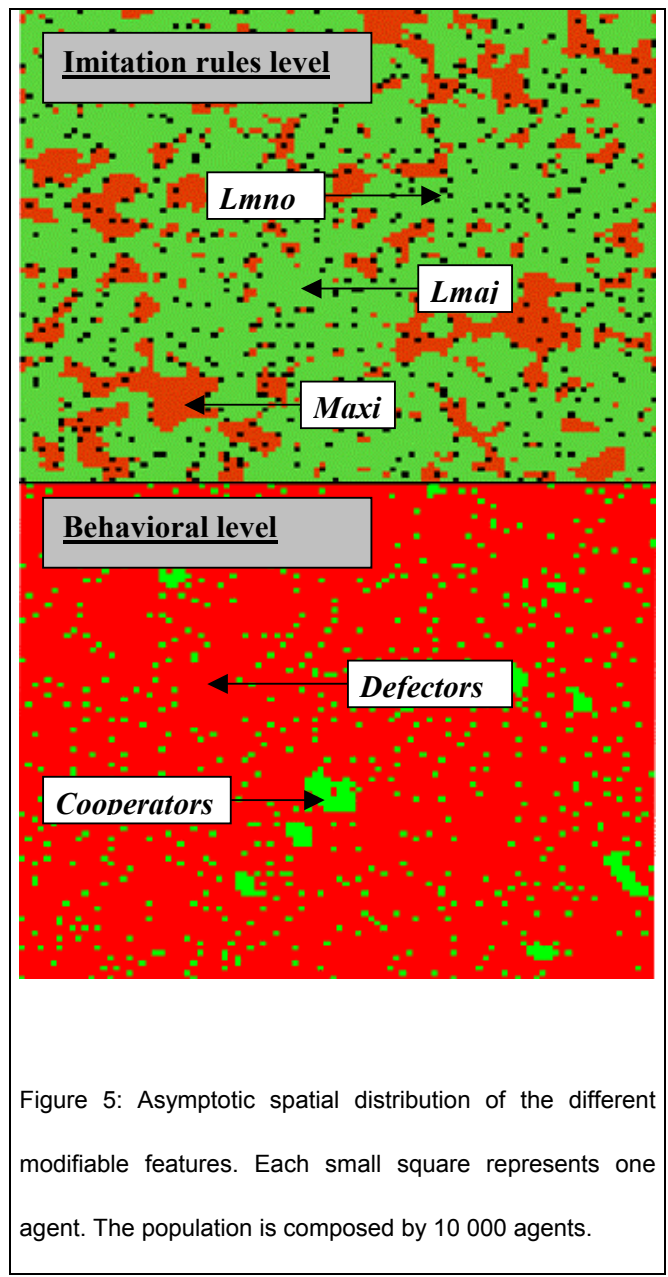

Figure 5: Asymptotic spatial distribution of the different modifiable features. Each small square represents one agent. The population is composed by 10 000 agents.

As for the behaviours, the asymptotic state is similar to what can be seen in figure 5, the size and number of clusters of cooperators being increasing with *Ci*. There is always a non null proportion of cooperators, between 5% (*b=2.5, Ci=0.1*) and 83% (*b=1.2, Ci=0.9*).

This kind of organization is not altered if the updates of the different modifiable features are asynchronous and these results are not artefacts of the parallel update. An interesting thing is the very weak variation of *Lmno* along all the parameter space studied. This suggests that the major factor for variations of *Lmno* proportions is the *topology of the network* more than the other state parameters. An interesting research would then be the study of the influence of topology on anti-conformist populations.

We begun to explore other settings of the model: continuous time, noisy updating and memory on past payoffs. The first experiments showed similar results: heterogeneous populations with strong attractors. The introduction of a memory on past payoffs increases the rate of cooperation in the population (We took for the payoffs $G_i$: $G_i^{t+1} = g(t) + \alpha G_i^t$, with $\alpha < 1$, and $g(t)$ being the scores of agent *i* at time *t*).

**Conclusions**

Observations of differences between human's and animal's meta-cognitive capacities has leaded us to propose some *cognitive hypotheses* on human mimetic processes. On this basis it is possible propose a framework for the modelling of human social systems which has the advantage of making mimetic processes endogenous and seems to have rich perspectives in explaining the emergence of structures in human societies. To give an example, we applied this formalism to a spatial game proposed by Nowak and May[23]. This allowed us to exhibit systems with stable self-regulated distribution of imitation rules and with heterogeneous distribution of behaviours and rules for imitation at the asymptotical states. Next step will be to apply this formalism to more complex models, and particularly to those dealing with the emergence of cooperation.

But this framework is also remarkable in the sense that it is specific to human's cognitive capacities. The explanations it could lead to, about evolution of some features of human societies, will consequently also be able to explain why this evolution has been so different from what have been observed in the animal world. From this point of view, meta-mimetism offers a way of thinking the relation between biological and cultural evolution. The fact that our genome has remained unchanged for several thousand years while our societies underwent the fundamental changes we know, suggests that the major part of evolution of social systems is no more played at the biological level but at the cultural level. Since mimetic processes operate on a daily basis when biological processes operate on a lifetime basis, they changed radically the path of selection, structuring human societies. For example, in the application presented here, cooperators naturally form clusters, which are known to be more robust to defecting behaviours than isolated cooperators. We had a chance animals never had: the possibility to evolve on much smaller time scales than the biological one through very particular mimetic behaviours. With the same body, we can change our cultural habits, begin new lives or even experiment several lives at the same time. The exploration/exploitation trade off that Nature is doing through mutation/selection processes on generations time scale has been transposed at the individual time scale through mimetic processes. The quick evolution of its entities is probably the real superiority of human social systems on animal systems.

**Methodology**

The algorithm for the simulations presented in this paper is the following:

*Set up of the game:*

- Give a value for *b* (here *1.2≤b≤2.5*).

- Neighbourhood composed by the eight adjacent cells plus the agent's cell.

*Initial Conditions:*

- Give the spatial distribution of imitation rules (here always uniform *i.i.d*)

- Give the spatial distribution of behaviours (here: *0.1≤ Initial rate of cooperators ≤0.9, i.i.d)*.

*At each period, for each agent:*

- The imitation rule is used to update itself. For example, if agent *A* had the *Lmaj* type and if the majority of her neighbors have turned to *Maxi* since last round, *A* will adopt the *Maxi* rule.

- The imitation rule (eventually new) is used to update the behavior. If *A*, a *Maxi* agent, played *C* last round but a *D*-player did strictly better than all *A*'s neighbors (*A* included), *A* will become a D-player.

- The agent plays with her nine neighbours (herself included).

- The new payoffs of agents are computed by summation of the nine scores of the two-players game.

**Supplementary Informations**

[1] For detailed presentation of the simulation presented here and example of simulations for the model without self-interaction see: http://chavalarias.free.fr/metamimetism.htm

**Acknowledgments :**

The authors thank T.K. Ahn, J.P. Dupuy, M.A. Janssen, N. Jonard, J. Petitot, L. Scubla, R.Topol, and G. Weisbuch for discussion and comments. This research is supported by the CNRS and MEL/OHLL.